# INVESTIGATION OF A BIOMASS GASIFICATION SYSTEM BASED ON ENERGY AND EXERGY ANALYSIS

Abbas Alpaslan Kocer[1], Yunus Emre Yuksel[2], Murat Ozturk[3]

[1] Uluborlu Selahattin Karasoy Vocational School, Suleyman Demirel University, 32260, Isparta Turkey, alpaslankocer@sdu.edu.tr
[2] Department of Science and Education, Education Faculty, Afyon Kocatepe University, 03200, Afyon, Turkey, yeyuksel@aku.edu.tr
[3] Department of Mechatronics Engineering, Faculty of Technology, Suleyman Demirel University, 32260, Isparta Turkey, muratozturk@sdu.edu.tr

**Abstract**

Biomass gasification procedure is a very complex process and it is influenced by many physical and chemical factors such as biomass gasification temperature and gasifier type. Thermodynamic assessment methodology based on the energy and exergy analysis can be used to evaluate the system performance and environmental impacts. In this paper, thermodynamic analysis of the biomass gasification system is given for the whole system and its components. The parametric studies reveal the effects of design and operating indicators on the exergy efficiency and exergy destruction rate. The result shows that the gasification temperatures for the biomass gasification system change significantly with the type of the gasifying medium.

## 1. Introduction

Energy is a key indicator for social, cultural and economic development of any country, and also is evaluated as an important aspect for sustainable development. It has been clearly seen that the energy production and consumption rate of a country is proportional to its economic status. By extension, the development of a country can be quantified as a ratio of its energy consumption per capita. Fossil energy sources, such as crude oil and natural gas, have been and refined to serve a dramatic growth in world population especially since the 1970s. Nevertheless, it is usually indicated that fossil energy sources are not sufficient to meet the constantly expanding needs of humanity. Conventional energy sources are non-renewable; they draw on finite sources that will finally dwindle, becoming more expensive or environmentally damaging to retrieve. Actually, at the nowadays consumption rate, conventional energy sources are reaching a natural discharge limitation with ongoing depletion [Ozturk et al., 2008]. Moreover, having relied merely on conventional energy sources has exhibited different significant environmental damages. Renewable energy sources are one of the most promising solutions for this energy demand. Alternative energy sources should preferentially be more environmentally and economical than conventional fossil energy sources in order to present wide scale applications. On the other hand, global warming, air pollution, acid precipitation, ozone depletion, forest destruction, and emission of radioactive substances are among the significant environmental problems [Ozturk et al., 2009]. Clean energy conversion and production variations with lower environmental concern should be obtained by considering all mentioned issues simultaneously [Dincer, 2000]. The usage of alternative energy sources provides a clean way to reduce the emissions of poisonous gases, such as $CO$, $CO_2$, $NO_x$ and $SO_x$. As an important example, in Turkey, approximately 25% of greenhouse gas emissions can be reduced by usage of renewable energy sources.

Many developed and developing countries installed intensive search plans in the before 1970s to install renewable energy technologies and change fossil energy sources [Ozturk et al., 2011]. The renewable energy technologies are the flat-plate solar panel installation for roofs of the residential and commercial building for heating and hot water production applications; photovoltaic (PV) system, wind turbine and ocean system for the electricity production; water splitting for hydrogen output; and biomass or bio-waste for conversion to gaseous fuel sources via gasification system for heat, steam or electricity generation.

Biomass is a large potential renewable energy sources, supplied from plants and animal wastes. It is one of the oldest renewable energy sources and has been used by humankind for daily needs since centuries [Toonssen et al., 2008]. To produce energy from biomass, the most preferred method is conventional combustion of biomass. This technique is not only valid in Turkey but also throughout the World. According to method which produces energy from biomass, biomass techniques are classified as classic and modern biomass. Classic biomass is the most popular method until now and this procedure consists of burning biomass such as wood, plant residues, and animal dung. Modern biomass technologies are new relative to classic ones and modern biomass is still on the development stages. In modern biomass technique, biomass is converted into solid, liquid or gas fuels by means of bio-chemical and thermo-chemical processes.

Investigations of the thermodynamic system are complex processes and involve consideration of the system components and their characteristic, chemical reaction and thermodynamic loss. Energy conversion technologies such as biomass gasification system should be investigated for their performance by using the first and second laws of thermodynamic (or energy and and exergy analysis). The use of exergy analysis should allow the determination of the processes having the greatest irreversibilities, as well as the causes and locations of the irreversibilities. Exergy analysis also would allow exergy efficiencies to be determined for whole system and its components. These important indicators should be used in design or retrofit of the process for increasing system performance. In this paper, energy and exergy analysis of the biomass gasification system and energy and exergy efficiencies of the system components are investigated for better system design. The simulations have been performed using Engineering Equation Solver (EES) software program. The following is a general outline of the present study;

- To investigate the effects of temperature and pressure of the gasifier, and biomass concentration variation on the biomass gasification system.
- To develop a theoretical model based on the thermodynamic laws performed in order to study in greater detail the effect of process flows on gasification system and investigate its performance.
- To calculate the exergy content for the system components including the chemical exergies for the biomass gasification plant.
- To investigate the performance assessments of the gasification system.

## 2. Properties of Biomass

Each biomass fuel has significantly different fuel properties, and the gasification chamber properties should be designed by using fuel properties. The specific properties of biomass sources are given in Table 1. Generally, biomass energy sources have less carbon, more oxygen and moisture contents than coal sources. Because of higher moisture and oxygen content, the lower heating values (LHV) and higher heating values (HHV) of biomass fuels are significantly lower than coal sources. The LHV and larger particle size of biomass sources cause storage difficulties. But, biomass sources have lower nitrogen and sulphur contents than coal sources, which are clearer, based on greenhouse gas emissions. The HHV of biomass sources is calculated from the following equation [Loo and J. Koppejan, 2008];

$$HHV_B = 0.3491 C_B + 1.1783 H_B + 1.005 S_B + 0.0151 N_B - 0.1034 O_B - 0.0211 A_B \qquad (1)$$

where subscript B is biomass fuel, C, H, S, N, O and A are the carbon, hydrogen, sulphur, nitrogen, oxygen and ash content of biomass sample in weight %, respectively. Proximate and ultimate analysis (wt%) of biomass examples are given in Table 2.

Table 1. Specific properties of biomass sources

| Specific properties | Values |
|---|---|
| Density (kg/m$^3$) | ~480-520 |
| Particle size (mm) | ~2.8-3.2 |
| SiO$_2$ Contents (wt% of dry ash) | ~23-49 |
| K$_2$O (wt% of dry ash) | ~4-48 |
| Fe$_2$O$_3$ (wt% of dry ash) | ~1.5-8.5 |
| Al$_2$O$_3$ (wt% of dry ash) | ~2.4-9.5 |
| Lower Heating Value (kJ/kg) | ~14,000-21,000 |

Table 2. Proximate and ultimate analyses (wt%) of biomass sources

|  | Proximate analysis (received basis) | | | | Ultimate analysis (dry ash free basis) | | | | | |
| --- | --- | --- | --- | --- | --- | --- | --- | --- | --- | --- |
|  | VM | FC | M | A | C | O | H | N | S | CI |
| Beech bark | 67.5 | 17 | 8.4 | 7.1 | 51 | 41.8 | 6 | 0.7 | 0.11 | 0.11 |
| Oak wood | 73 | 20 | 6.5 | 0.3 | 50 | 42.9 | 6.1 | 0.3 | 0.10 | - |
| Sawdust | 55.1 | 9.3 | 34.9 | 0.7 | 49 | 43.4 | 6.1 | 0.7 | 0.11 | 0.01 |
| Switch Grass | 70.8 | 12.8 | 11.9 | 4.5 | 49 | 43.4 | 6.1 | 0.7 | 0.11 | 0.08 |
| Straw | 64.3 | 13.8 | 12.4 | 9.5 | 48 | 44.5 | 5.6 | 1 | 0.13 | 0.54 |
| Almond shell | 69.5 | 20.2 | 7.2 | 3.1 | 50 | 42.5 | 6.2 | 1 | 0.05 | 0.06 |

## 3. System Design

Figure 1 shows a schematic of the biomass gasification system, modeled for the theoretical investigation. Both air and the biomass fuels enter the gasifier at the environment temperature and pressure. Gasification takes place in the gasification chamber and the flue gases after exchanging the heat with the feed water exit through the stack at 155 °C. The most of the ash, which is assumed approximately as 80% fly ash, exits the gasifier with the flue gases by the chimney. Gasification technology converts biomass fuels into product gases that should be used in energy conversation technology as an input gas fuel. The product gases are generally consists of CO, $CO_2$, $CH_4$ and $H_2$, and combusted to generate heat and shaft work. The produced gasses are also used as feedstock for the production of synthesis gas, liquid fuels and different chemicals. The lower heating values of the produced gases should be determined via the gasses composition data. The outputs of the biomass gasification chamber also include unused materials, such as particulars, tar, ammonia and hydrogen sulfide.

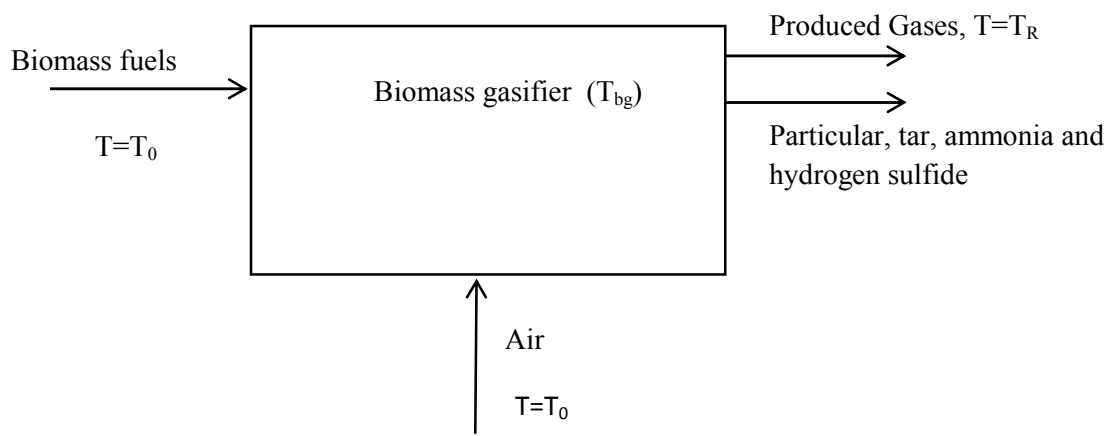

Figure 1. Schematic diagram of the biomass gasification system

The biomass gasification reaction for 1 kg dry biomass source should be written as follows;

$$(cC + hH + oO + nN + sS) + v_{H_2O} H_2O = v'_{CO_2} CO_2 + v'_{CO} CO + v'_{H_2} H_2 + v'_{H_2S} H_2S + v'_{N_2} N_2 + v'_{H_2O} H_2O + v'_{CH_4} CH_4 \qquad (2)$$

Here, c, h, o and n values are used as kmol/kg, and mass fraction.

## 4. Thermodynamic Analysis

Thermodynamic analysis of a biomass gasification system and its components requires special consideration of multiple aspects, such as fluid dynamics, gasifier design, gas-dynamics and thermodynamics of expanding flows with non-linear behavior and heat transfer. The proposed mathematical model for a biomass gasification system includes the conservation of mass, energy and exergy. To mathematically describe the biomass gasification system, a comprehensive knowledge of physical and chemical exergy content for processes is required. Mass, energy, entropy and exergy balance equations should be written for system components and for the whole system under the assumption of steady state operation. This is done in regard to a selected working fluid in a given cycle configuration and imposed operating conditions of source. A general balance equation for any quantity in a system can be written as [Dincer and Rosen, 2013];

*Input + Generation – Output - Consumption = Accumulation*

The general mass, energy, entropy and exergy balance equations for system thermodynamic analysis on control volumes of process components are written as follows, respectively;

Mass balance equation;

$$\sum \dot{m}_{in} = \sum \dot{m}_{out} \tag{3}$$

Energy balance equation;

$$\sum \dot{m}_{in} h_{in} + \sum \dot{Q}_{in} + \sum \dot{W}_{in} = \sum \dot{m}_{out} h_{out} + \sum \dot{Q}_{out} + \sum \dot{W}_{out} \tag{4}$$

Entropy balance equation;

$$\sum \dot{m}_{in} s_{in} + \sum \left(\frac{\dot{Q}}{T}\right)_{in} + \dot{S}_{gen} = \sum \dot{m}_{out} s_{out} + \sum \left(\frac{\dot{Q}}{T}\right)_{out} \tag{5}$$

where T is the temperature at where heat transfer cross the system boundary. Exergy balance equation;

$$\sum \dot{m}_{in} ex_{in} + \sum \dot{W}_{in} = \sum \dot{m}_{out} ex_{out} + \sum \dot{W}_{out} + \dot{Ex}_D \tag{6}$$

where $\dot{Ex}_D$ is the exergy destruction rate. Exergy analysis should be given as the highest content of work that should be derived by investigating processes that bring the system into equilibrium [Szargut et al., 1988; Rosen, 1986].

### 4.1 General efficiency equations

The energy efficiency ($\eta$) of the investigated system should be given as the ratio of useful energy produced by the process to the total energy input. The useful produced energy represents the desired results produced by the system components, such as electricity, heating and cooling, hot water, hydrogen and other chemicals. The energy efficiency for steady-state processes should be written as follows:

$$\eta = \frac{useful\ energy\ output\ rate\ with\ products}{total\ energy\ input\ rate} = \frac{\dot{E}_{out,useful}}{\sum \dot{E}_{in,total}} \qquad (7)$$

Exergy efficiency of the system components and whole system give the main effectiveness of each process of the system. The exergy efficiency ($\psi$) of the investigated system can be given as the divided of exergy output rate ($\dot{E}x_{out,useful}$) that is created by the considered system to the overall exergy inlet rate ($\sum \dot{E}x_{in,total}$) that is cross the boundaries of the system. The exergy efficiency for steady-state processes should be written as follows:

$$\psi = \frac{total\ useful\ exergy\ output\ rate\ with\ products}{total\ exergy\ input\ rate} = \frac{\sum \dot{E}x_{out,total}}{\sum \dot{E}x_{in,total}} \qquad (8)$$

### 5. Assumptions

The chosen assumptions for this paper are given as follows:
- All the system components and whole system operate at steady state conditions.
- All the proses gases are considered as ideal gases.
- References ambient temperature and pressure are chosen as 25 °C and 1 atm, respectively.
- Ambient reference air considerations are considered as 21% oxygen and 79% nitrogen on the volume basis.
- Heat loss via radiation and convection from the gasifier to the environment is 1-2% of fuel energy input [Basu et al., 2000; Fauklker and de Saouza-Santos, 2010].
- Kinetic and potential energy impacts are neglected.

### 6. System Analysis

The exergy contents of flowing material have two indicators, such as physical exergy part and chemical exergy part. In the gasification system, both physical and chemical exergy indicators should be required because chemical reactions take place in the gasifier while only physical exergy is combined by the steam process elements. The specific exergy for an investigated state should be written as follows;

$$ex = ex_{ph} + ex_{ch} \qquad (9)$$

The physical exergy or specific flow exergy should be given as follows;

$$ex_{ph} = (h - h_o) - T_o(s - s_o) \qquad (10)$$

The chemical exergy contents of ideal gas should be given as follows;

$$ex_{ch} = ex_{ch}^o + R_u T_o \ln z_i \qquad (11)$$

where, $z_i$ is the mole fractions of the i[th] components and $ex_{ch}^o$ is the molar chemical exergy at the given reference temperature and pressure, and should be given as follows [Kotas, 1980];

$$ex_{ch}^o = \frac{T_o}{T} ex_{o,i}^{ch} - \bar{h}_f^o \frac{T - T_o}{T} \qquad (12)$$

where $ex_{o,i}^{ch}$, $\bar{h}_f^o$ and T are standard molar chemical exergy of i$^{th}$ chemical composition, formation enthalpy and gasifier temperature, respectively. The specific values of standard molar chemical exergies of the various substances given in the theoretical investigation are given in Table 3 [Cengel and Boles, 2006].

Table 3. Standard enthalpy and chemical exergy of various substances

| Substance | Standard enthalpy (MJ/mol) | Standard chemical exergy (MJ/mol) |
|---|---|---|
| $O_2$ (g) | 0 | 3.97 |
| $N_2$ (g) | 0 | 0.72 |
| $CO_2$ (g) | -393.52 | 19.87 |
| $H_2O$ (g) | -241.82 | 9.5 |
| $H_2O$ (l) | -285.83 | 0.9 |
| $SO_2$ (g) | -297.10 | 313.40 |
| NO (g) | 90.59 | 88.90 |
| $NO_2$ (g) | 33.72 | 55.60 |

The exergy balance equation of the gasification chamber should be given as follows;

$$\dot{E}x_{bf} + \dot{E}x_a + \dot{E}x_Q = \sum_e \dot{E}x_{hg} + \dot{E}x_D^G \tag{13}$$

where $\dot{E}x_{bf}$, $\dot{E}x_a$, $\dot{E}x_Q$, $\dot{E}x_{hg}$ and $\dot{E}x_D^G$ are exergy rate of biomass fuels, air, heat, hot product gases and exergy destruction rate of gasifier, respectively.

Biomass fuels and air enter the gasifier at reference temperature and pressure, therefore their physical exergies equal zero. In this paper, specific chemical exergies of biomass fuels are calculated using the method which proposed by Szargut [2005], and this methology should be given as follows;

$$ex_{bf}^{ch} = \beta_{bf}(LHV_{bf} + wh_{hg}) + wex_{ch,w}^o \tag{14}$$

where w is weight percent (w%) of moisture in fuel and $ex_{ch,w}^o$ is the specific chemical exergy of water at ambient temperature and pressure. The coefficient $\beta_f$ is derived from experimental results, and should be calculated as follows;

$$\beta_{bf} = \frac{1.044 + 0.016(H/C) - 0.3493(O/C)[1 + 0.0531(H/C)]}{1 - 0.4124(O/C)} \tag{15}$$

The chemical exergy of the air should be determined as follows;

$$\dot{E}x_a = A_a[\xi_{O_2}^o + 3.76\xi_{N_2}^o + R_u T_o\{ln(0.21) + ln(0.79)\}] \tag{16}$$

where $A_a$ is air molar flow rate, $\xi_{O_2}^o$ and $\xi_{N_2}^o$ are specific chemical exergy of oxygen and nitrogen, respectively.

## 7. Results and Discussion

In this paper, Engineering Equation Solver (EES) software program utilized to investigate the modeling biomass gasification system based on the thermodynamic assessment methodology.

Also, EES contains built in library of thermodynamic data for many chemical substances. Exergy destruction rate, energy efficiency and exergy efficiency of the biomass gasification system components are investigated in the present paper, besides the assessment of system components performance. Heat is lost from the gasifier to the environment during the process that has been approved in many papers. Heat transfer through the temperature difference always increases the entropy generation or exergy destruction rate. In order to investigate the gasifier efficiency, analysis results of exergy destruction rate can help to identify the defect within the gasification plant, and the system efficiency should be improved via modification of the biomass gasification plant from considering reduction of heat losses in the further system design.

Energy contents of matters are the measure of quantity, but exergy contents of matters are measure of both quantity and quality of energy contents. Exergy values of biomass fuels are approximately 20% greater than energy values by exergetic indicator (β). In addition to that, exergetic indicators of biomass fuels are higher than coal samples. The system performances have been investigated in terms of key indicators. Gasifier temperature, ambient temperature and pressure, work output, exergy destruction rate and main losses are used for efficiency analysis of the biomass gasification system. It is also observed that moisture contents of biomass fuels have more important effects on the decrease of gasifier outlet temperatures than ash contents.

The effects of varying ambient temperature from 10 °C to 30 °C on the exergy destruction rate and exergy efficiency of the biomass gasification system are given in Figure 2. According to the Figure 2, the exergy destruction rate of the biomass gasification system decreases with rising ambient temperature. On the other hand, exergy efficiency of the biomass gasification system increases with increasing ambient temperature.

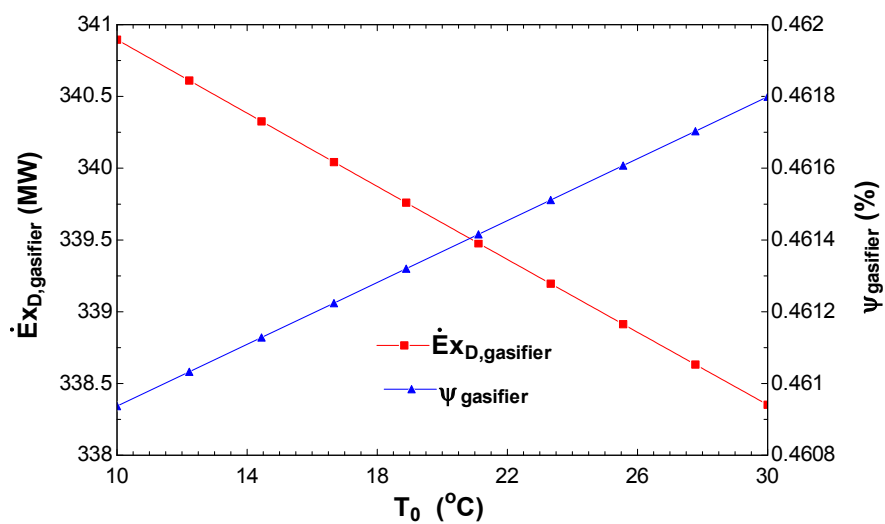

Figure 2. Variations with ambient temperature of the exergy destruction rate and exergy efficiency for the biomass gasification system

Figure 3 shows that, the exergy destruction rate of the gasifier decreases with increasing gasifer temperature from 625 °C to 850 °C, but its exergy efficiency increases. The variations of exergy destruction rate and exergy efficiencies of gasifer remain almost linear depending on the ambient and gasifier temperature. These results are expected since the exergy destruction rate and exergy efficiency of the process are usually inversely proportional properties.

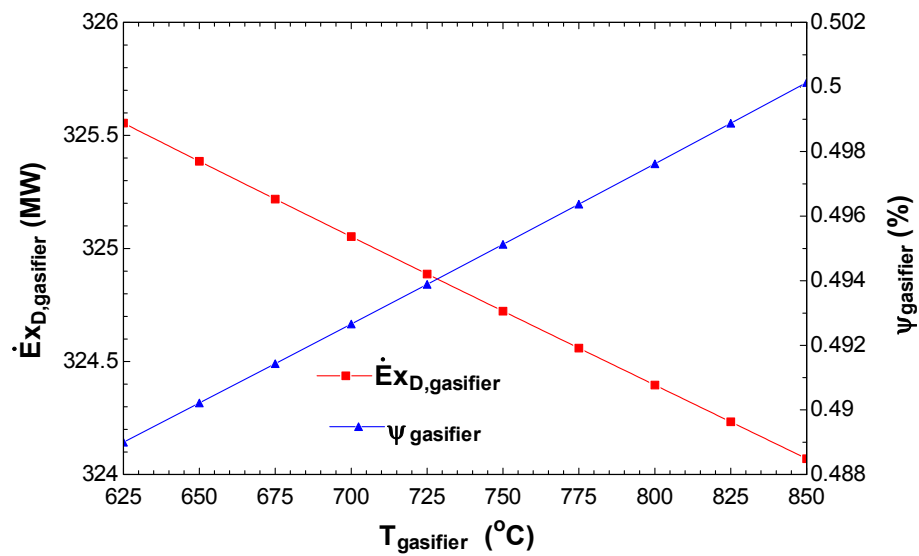

Figure 3. Variations with gasifier temperature of the exergy destruction rate and exergy efficiency for the biomass gasification system

## 8. Conclusions

The main objective of the present study is to investigate the biomass gasification system operating characteristics that provide power for various operating conditions by using parametric analyses. The parametric analyses compare the system components and their responses to variations in selected operating conditions, such as ambient temperature and gasifer temperature. In addition to that, exergy efficiency of the biomass gasification is investigated to indicate how the gasification process reaches the real operating conditions. The main conclusions of the present paper should be given as follows;

- Mass, energy and exergy balance equations for the system components and whole system are necessary to investigate the gasification system performance.
- The parametric studies are very useful to investigate the variations of system efficiency for changing operating indicators.
- Decreasing the exergy destruction rate of biomass gasification system components and whole system, and increasing the energy and exergy efficiency result in decreased greenhouse gas emissions, less environmental impacts and increased sustainability.
- Important cause for higher exergy destruction rate or lower exergy efficiency in the present paper is highly irreversible chemical reaction in the biomass gasifier.